\documentclass[10pt,conference]{IEEEtran} 

\usepackage{algpseudocode}
\usepackage{amsmath}
\usepackage{array}
\usepackage{balance}
\usepackage{booktabs}
\usepackage[font=small,labelfont=bf]{caption}
\usepackage{hyperref}
\usepackage[capitalise]{cleveref}
\usepackage{color}
\usepackage{colortbl}
\usepackage{enumitem}
\usepackage{graphicx}
\usepackage{listings}
\usepackage{mdframed}
\usepackage{MnSymbol}
\usepackage{multirow}
\usepackage{soul}
\usepackage{subcaption}
\usepackage{tabularx}
\usepackage{tcolorbox}
\usepackage{titlesec}
\usepackage{xurl}
\usepackage{wrapfig}
\usepackage{xcolor}
\usepackage{listings}
\usepackage{xcolor}
\usepackage{tabularx}
\usepackage{xspace}

\definecolor{diffred}{RGB}{220,50,47}
\definecolor{diffgreen}{RGB}{38,139,210}

\lstdefinelanguage{Diff}{
  morecomment=[f][\color{diffred}]{-},
  morecomment=[f][\color{diffgreen}]{+},
  morecomment=[f][\color{gray}]{@},
  morecomment=[f][\color{gray}]{@@}
}

\definecolor{latex_green}{HTML}{4c886b}
\definecolor{latex_cyan}{HTML}{318495}
\definecolor{latex_blue}{HTML}{0000ff}
\definecolor{latex_gray}{HTML}{646464}
\definecolor{latex_purple}{HTML}{7d2884}

\lstdefinestyle{lststyle}{
    commentstyle=\color{latex_green},
    keywordstyle=\color{latex_blue},
    numberstyle=\tiny\color{latex_gray},
    stringstyle=\color{latex_purple},
    basicstyle=\ttfamily\footnotesize,
    breakatwhitespace=false,         
    breaklines=true,                 
    captionpos=b,                    
    keepspaces=true,                 
    numbers=left,                    
    numbersep=5pt,                  
    showspaces=false,                
    showstringspaces=false,
    showtabs=false,                  
    tabsize=2
}
\newcommand{\tool}{\texttt{VRRefMiner}}
\newcommand{\toolminer}{\texttt{VRRefMiner}\xspace}

\AtBeginDocument{%
  }

\definecolor{promptcolor}{HTML}{2D4671}
\definecolor{devcolor}{HTML}{805A15}

\definecolor{rqs}{HTML}{aa5239}

\definecolor{findings}{HTML}{28784D}

\definecolor{hlcolora}{HTML}{FFFF66}

\definecolor{hlcolor}{HTML}{83C0A0}

\title{Mining Unity VR Repositories for Refactoring Patterns and Technical Debt}

\begin{document}


\author{%
\IEEEauthorblockN{Faraz Gurramkonda\IEEEauthorrefmark{1},
Neeraj RandhirSingh Saini\IEEEauthorrefmark{1},
Michael Gharbieh\IEEEauthorrefmark{1},\\
Bruce Maxim\IEEEauthorrefmark{1},
M. Rasel Mahmud\IEEEauthorrefmark{2},
and Foyzul Hassan\IEEEauthorrefmark{1}}
\IEEEauthorblockA{\IEEEauthorrefmark{1}University of Michigan-Dearborn}
\IEEEauthorblockA{\IEEEauthorrefmark{2}Kennesaw State University}
}
\maketitle


\begin{abstract}
Maintaining virtual reality (VR) applications poses challenges tied to real-time performance, hardware dependencies, and complex interactions. Yet little is known about how developers refactor these applications, which kinds of technical debt they address, or how well automated tools capture VR-specific changes. We mined 1,755 candidate refactoring commits from 176 open-source Unity VR repositories and manually examined a sample of 303 commits, of which 270 contained at least one refactoring. We also extended RefactoringMiner to create VRRefMiner, a tool that detects conventional refactorings in C\# scripts used by Unity applications.

Our analysis found both conventional and VR-specific refactorings. VR-specific changes addressed concerns such as interaction logic, performance, and Unity-specific artifacts. Code debt was the most prevalent category overall, while interaction debt and UI/UX debt were the main categories addressed by VR-specific refactorings. Across the repositories, VRRefMiner detected 15,549 operations spanning 68 refactoring types, with structural moves and extractions accounting for 31.0\% of detections. It detected conventional refactorings in 216 of 246 commits labeled as general refactoring. In the 24 commits labeled as VR-specific, the tool could detect conventional changes but identified no VR-specific refactoring categories. These results point to a need for detection techniques that account for both source code and the Unity artifacts involved in VR maintenance.

\end{abstract}

\setlength{\textfloatsep}{1pt}
\setlength{\abovecaptionskip}{1pt}
\setlength{\belowcaptionskip}{1pt}
\setlist{nolistsep}

\section{Introduction}
\label{sec:introduction}
Virtual Reality (VR) software is transforming how users interact with digital environments across diverse domains, including gaming, education, healthcare, and industrial training. The global VR market was valued at USD 15.9 billion in 2024 and is projected to grow from USD 20.48 billion in 2025 to USD 38.0 billion by 2029~\cite{vrmarket}. This rapid growth reflects the increasing adoption of VR technologies and underscores the importance of developing reliable and maintainable VR software systems. Unlike traditional software, VR applications face distinct software engineering challenges arising from stringent real-time performance requirements, complex interaction logic, and tight coupling with specialized hardware platforms~\cite{rzig2023vrtesting,lin2019unity,amershi2019interactive}. These characteristics introduce additional complexities in software development, maintenance, and evolution, highlighting the need for effective software engineering practices tailored to VR systems~\cite{zhuvrexplorer,rzig2023vrtesting,Wang2022_VRTest,Rafi2022_PredART}.

Refactoring is a fundamental software maintenance practice that improves a software system’s internal structure without changing its observable behavior~\cite{wahler2016improving,oliveira2023untold,palomba2017exploratory}. Prior research has shown that refactoring can help reduce technical debt and improve code readability, maintainability, and, in certain contexts, performance~\cite{peruma2022refactoring,traini2021software}. Researchers have also developed automated techniques for detecting and recommending refactorings, using machine learning, mining software repositories, and static analysis to identify refactoring opportunities and evaluate their impact~\cite{hu2019re,meng2015does,ouni2016multi}. In addition to traditional software systems, prior studies have investigated refactoring practices and automated approaches in specialized domains, including machine learning~\cite{noei2025detecting,tang2021empirical}, Android applications~\cite{fawad2025refactoring,lin2015study}, and DevOps~\cite{ksontini2025refactoring,ksontini2021refactorings,ghammam2025build}.

However, despite substantial progress in refactoring research across diverse software domains, refactoring practices in VR software systems remain largely unexplored.  Prior studies have highlighted the unique software maintenance challenges of VR applications, including performance optimization, testing, and limitations of existing automated analysis techniques in addressing VR-specific quality concerns~\cite{Nusrat2021,rzig2023vrtesting, WangASE2023, Gurramkonda2025,Zhu2025}. VR applications integrate source code with domain-specific artifacts, including scenes, game objects, and components, which may introduce refactoring activities beyond conventional source-code transformations. Although existing automated refactoring tools, such as RefactoringMiner~\cite{tsantalis2020refactoringminer}, can identify various structural code changes, their effectiveness in detecting VR-specific refactorings remains unclear. Furthermore, limited empirical evidence exists on the types of refactorings performed in VR projects and the technical debt these changes aim to address. These gaps highlight the need for empirical research to better understand VR refactoring practices and assess the applicability of existing automated detection techniques to immersive software systems.

To address these research gaps, we conduct an empirical study of refactoring practices in open-source VR software systems, focusing on applications developed using Unity and C\#. Specifically, we investigate the following research questions:

\begin{itemize}
\item \textbf{RQ1: What types of refactorings are commonly performed in VR projects?}
\item \textbf{RQ2: What types of technical debt do these refactorings aim to address?}
\item \textbf{RQ3: How effectively do existing automated refactoring detection tools identify refactorings in VR projects?}
\end{itemize}

To answer these research questions, we employ a mixed-methods research approach, as detailed in~\autoref{sec:approach}. We analyze a dataset of 314 open-source VR applications curated by Rzig et al.~\cite{rzig2023vrtesting}, from which we identify 1,755 potential refactoring commits. We then conduct manual validation and qualitative analysis on a sample of 303 commits to characterize refactoring types and identify the technical debt they aim to address. Furthermore, to investigate RQ3, we extend RefactoringMiner to develop \tool{} (~\autoref{subsec:tool}), a tool that supports automated detection of conventional refactorings in C\# code from Unity-based VR applications. We focused on Unity due to its widespread use in VR development. A recent study reported that 82\% of surveyed XR applications developed between 2016 and 2020 used Unity~\cite{xr_development_2026}.

Our empirical evaluation (~\autoref{sec:evaluation}) reveals that conventional refactorings are prevalent in VR projects, while VR-specific refactorings also occur, particularly those involving interaction logic, input handling, and performance-related subsystems (~\autoref{subsec:rq1_results}). Regarding technical debt, code debt is the most prevalent category overall, whereas VR interaction debt and VR UI/UX debt are the primary types addressed by VR-specific refactorings (~\autoref{subsec:rq2_results}). Furthermore, our tool (\tool{}) identified 15,549 conventional refactoring operations across 68 distinct types, with structural moves and extractions accounting for 31.0\% of detections. In our manually labeled dataset, \tool{} detected conventional refactorings in 216 of 246 general-refactoring commits (87.80\%) but did not identify VR-specific refactoring categories in any of the 24 VR-specific commits (~\autoref{subsec:rq3_results}). These findings highlight the distinct maintenance needs of VR applications and the importance of developing VR-aware refactoring detection techniques.


Overall, our main contributions are as follows:

\begin{itemize}
    \item \textbf{VR Refactoring Dataset:} We construct a dataset of 1,755 potential refactoring commits from 176 open-source VR repositories (~\autoref{subsec:data}), providing a foundation for future empirical studies and automated refactoring techniques tailored to VR software.

    \item \textbf{Empirical Characterization of VR Refactoring:} We conduct a quantitative and qualitative study of refactoring practices in Unity-based VR projects. We develop a taxonomy encompassing both conventional and VR-specific refactoring types (\autoref{subsec:rq1_results}) and investigate their relationships with different categories of technical debt (\autoref{subsec:rq2_results}), revealing maintenance concerns unique to immersive software systems.

    \item \textbf{Automated Refactoring Detection and Evaluation:} We develop \tool{} (\autoref{subsec:tool}) by extending RefactoringMiner to detect conventional refactorings in Unity C\# code. We systematically evaluate its effectiveness, identifying 15,549 refactoring operations across 68 types and revealing limitations in detecting VR-specific refactoring activities (\autoref{subsec:rq3_results}).

\end{itemize}

Our replication package is available on figshare~\cite{verefactoring}. 
\vspace{-0.2cm}
\section{Related Work}
\label{sec:related}


\subsection{Refactoring in Traditional and Domain-Specific Software}
Refactoring improves software structure while preserving external behavior. Prior empirical studies have investigated how developers perform refactoring, their motivations, and its impact on software quality. Murphy-Hill et al.~\cite{Murphy-Hill2012} examined how developers perform refactoring in practice, while Tsantalis et al.~\cite{Tsantalis2013} conducted a multidimensional empirical study of refactoring activities. Beyond traditional software, researchers have investigated refactoring in specialized domains. Tang et al.~\cite{tang2021empirical} identified 14 ML-specific refactoring categories, while Habchi et al.~\cite{HABCHI2021110964} examined the introduction and removal of Android-specific code smells. Similarly, Ksontini et al.~\cite{ksontini2021refactorings} identified 24 Docker-specific refactoring types, and Ghammam et al.~\cite{ghammam2025build} characterized 24 build-script refactoring types across six categories. Recent work also investigated Dockerfile refactoring practices and opportunities for automation~\cite{Ksontini2025}. Giannaccari et al.~\cite{giannaccari2025code} investigated the use of virtual reality as an interface for performing source-code refactorings. In contrast, our study examines refactoring activities within the source code and related components of VR applications, focusing on their types, associated technical debt, and automated detection. Prior research has also investigated code smells and maintainability issues in video games and Unity-based applications. Borrelli et al. ~\cite{borrelli2020unity} and Nardone et al.~\cite{nardone2023videogame} examined maintainability concerns in game software, while Bosco et al.~\cite{bosco2023unitylint} introduced UnityLint to identify Unity-specific code smells. However, code smells and refactorings are distinct concepts: code smells indicate potential design or maintainability problems, whereas refactorings are changes to the internal structure of software intended to preserve its external behavior. Unlike code-smell detection studies, our work investigates refactoring operations performed by developers in VR software projects.These studies demonstrate that refactoring practices can vary across software domains. However, refactoring in VR software remains underexplored, particularly given the interactions among source code, scenes, assets, and runtime behavior. Our study addresses this gap through an empirical investigation of VR refactoring practices.

\subsection{Software Engineering Practices in VR Systems}
Prior research has investigated software engineering challenges in Virtual Reality (VR) applications, particularly performance optimization, automated testing, and software quality. Nusrat et al. examined optimization practices in Unity-based VR projects, highlighting maintenance activities driven by real-time performance requirements~\cite{Nusrat2021}. Other studies have investigated testing practices and automated testing techniques for VR applications, emphasizing challenges associated with scene structures, user interactions, and runtime behaviors~\cite{rzig2023vrtesting,Wang2022_VRTest}. More recent studies, including VRTestSniffer~\cite{Gurramkonda2025} and VRExplorer~\cite{Zhu2025}, have further examined VR-specific testing issues and techniques for analyzing immersive software systems. Despite these advances, existing research has primarily focused on performance optimization and testing, leaving refactoring practices in VR software insufficiently explored. In particular, limited empirical evidence exists on VR-specific refactoring types, their relationships with technical debt, and the effectiveness of automated refactoring detection techniques. Our study addresses these gaps through an empirical investigation of refactoring practices in Unity-based VR projects.

\section{Research Approach}
\label{sec:approach}
\begin{figure*}[t]
    \centering
    \includegraphics[width=\linewidth]
    {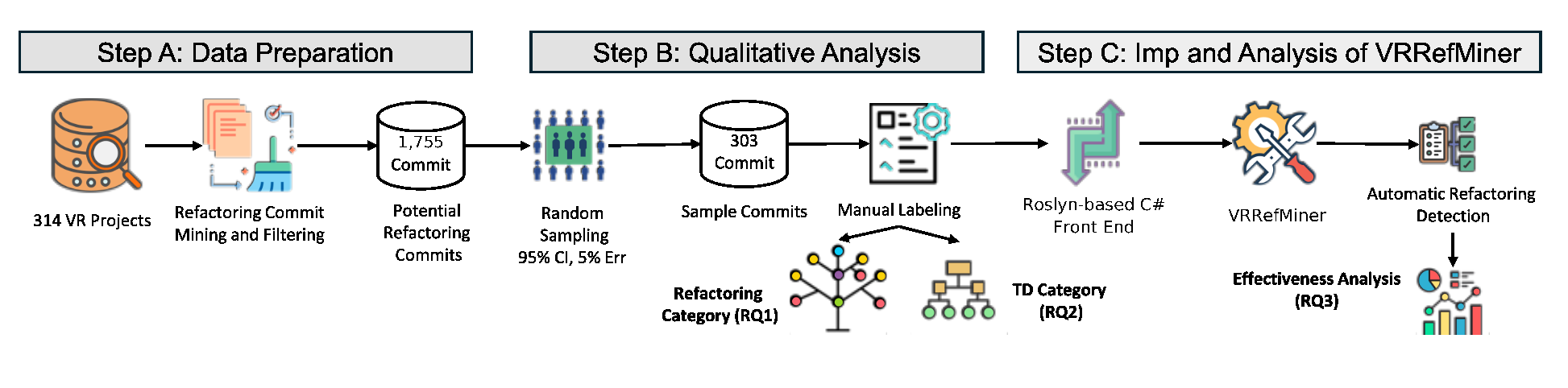}
    \vspace{-0.7cm}
    \caption{Overview of the Research Approach.}
    \vspace{-0.6cm}
    \label{fig:approach-overview}
\end{figure*}

~\autoref{fig:approach-overview} illustrates our research methodology, which consists of three main phases: (1) Data Preparation (~\autoref{subsec:data}), (2) Refactoring and Technical Debt Analysis (~\autoref{subsec:manual}), and (3) Automated Refactoring Detection using \tool{} (~\autoref{subsec:tool}).



\subsection{Data Preparation}
\label{subsec:data}
\vspace{-0.1cm}
For our refactoring analysis, we used a curated dataset of 314 manually verified Unity-based VR projects, originally compiled in prior work by Rzig et al.~\cite{rzig2023vrtesting}. These projects encompass a diverse range of Unity-based VR applications, including 164 independent, 67 organizational, and 83 academic projects, providing a heterogeneous dataset for investigating refactoring practices in open-source VR software. To identify refactoring-related commits, we performed commit mining using JGit to analyze the complete commit history of each Unity-based VR project in our dataset. We employed a keyword-based filtering mechanism, selecting commits whose messages contained ‘refactor*’ to capture common variants such as ‘refactors’, ‘refactoring’, and ‘refactored’. This technique, widely adopted in prior refactoring research~\cite{alomar2021refactoring,oliveira2023untold,palomba2017exploratory,tang2021mltdrefactoring,ksontini2021refactorings}, efficiently narrows the search space to code changes explicitly documented as refactoring. The choice of ‘refactor*’ was intentionally made to reduce false positives~\cite{traini2021software,hu2019re} and increase the likelihood of identifying genuine refactoring activities. Additionally, our mining process focused on Unity-specific files and directories—including \textit{.cs} scripts and directories such as \textit{Assets/} and \textit{Scenes/}—to ensure relevance to VR application development. To improve the precision of our refactoring commit selection, we excluded commits modifying more than 20 source files. Large commits are more likely to contain multiple interleaved changes, commonly referred to as tangled changes~\cite{Herzig2013,Azad2023,Stana2025}, which can complicate the identification and classification of individual refactoring activities. Excluding such commits helps reduce noise and facilitates reliable manual analysis. The 20-file threshold is informed by prior software engineering research, including studies showing that 89\% of bugs can be localized within 20 files~\cite{Chen2021}, as well as studies on bug analysis~\cite{Khan2022} and bug prediction~\cite{Lewis2013}. Finally, we removed duplicate commits, two commits that were no longer accessible, and merge commits to avoid redundancy and ensure the availability of changes for analysis. After applying these filtering criteria, we obtained \textbf{1,755 unique commits} from \textbf{176 distinct VR repositories}. This dataset serves as the foundation for both large-scale characterization and tool-based analysis, as explored in our empirical evaluation.

\subsection{Qualitative Analysis}
\label{subsec:manual}

Given the substantial size of our dataset and the effort required for manual examination, we employed random sampling to select a representative subset of potential refactoring commits. Sampling is a widely adopted practice in qualitative analysis and empirical software engineering, particularly when analyzing large datasets~\cite{levin2019towards,rzig2023vrtesting,LiASE2026,Dintzner2016}. Since refactoring practices in VR software remain largely unexplored, manual examination was essential to identify conventional and VR-specific refactoring activities (RQ1,~\autoref{subsec:rq1_results}) and investigate their relationships with different types of technical debt (TD) in VR development (RQ2, ~\autoref{subsec:rq2_results}).


We employed an open-coding approach to identify refactoring activities without imposing predefined categories, allowing both conventional and VR-specific refactoring types to emerge from the data, following practices adopted in prior studies of domain-specific refactoring~\cite{ghammam2025build}. 

Guided by the taxonomy development framework of Usman et al.~\cite{usman2017taxonomies}, two co-authors with expertise in VR and refactoring (3 years experience in C\#, Unity, VR development, and refactoring analysis) independently examined the sampled commits in two rounds. The identified refactorings were then classified into general software engineering or VR-specific categories based on the observed transformations. In the first round (Identification), labelers identified refactoring changes and explicit evidence of technical debt (TD) using commit messages, code diffs, VR framework documentation, and the behavior-preserving definition of refactoring. Non-refactoring commits were excluded from subsequent analysis. In the second round (Categorization), they classified the identified refactorings and associated TD, linking refactorings to TD categories only when supported by explicit evidence. The disagreements between annotators were reviewed through discussion, with the researchers examining the corresponding code changes and commit context before reaching a consensus. The agreed-upon labels were then used to finalize the refactoring taxonomy, yielding a Cohen's $\kappa = 0.923$ for RQ1 (refactoring classification) and an agreement rate of 82.13\% for RQ2 (technical debt classification). Finally, we employed card sorting~\cite{fincher2005making,spencer2004card} to organize the identified refactoring types into a hierarchical taxonomy of categories and subcategories.

Commit messages containing ``refactor'' identified potential refactoring commits, while manual inspection determined their validity. Changes limited to comments or altering runtime behavior without structural refactoring were excluded.

\subsection{Automated Refactoring Detection with \tool{}}
\label{subsec:tool}
State-of-the-art refactoring detection tools, such as RefactoringMiner~\cite{tsantalis2018accurate}, identify refactoring operations by analyzing structural changes between source-code revisions, typically using abstract syntax trees (ASTs). Although their underlying detection principles are applicable across programming languages, implementations often depend on language-specific parsing and analysis capabilities. Existing tools, including RefDiff~\cite{Silva2017} and RefDetect~\cite{Moghadam2021}, support languages such as Java, JavaScript, C, and C++, but lack C\# support. Recent versions of RefactoringMiner have expanded support beyond Java to Python, Kotlin, JavaScript, TypeScript, and C++; however, its latest release (v3.1.6) still does not support C\#. While SemanticDiff~\cite{sysmagine2026semanticdiff} offers limited C\# refactoring detection and RefactoringMinerExtension~\cite{kalaitzidis2025refactoringminer} explores extending RefactoringMiner to additional languages, comprehensive refactoring detection for real-world Unity projects remains limited. Furthermore, static analysis tools such as SonarQube~\cite{sonarsource2025sonarqube} and JetBrains InspectCode~\cite{jetbrains2026inspectcode} primarily identify code quality issues rather than retrospectively detecting refactoring operations between revisions.

\begin{figure}[t]
    \centering
    \includegraphics[width=\columnwidth]{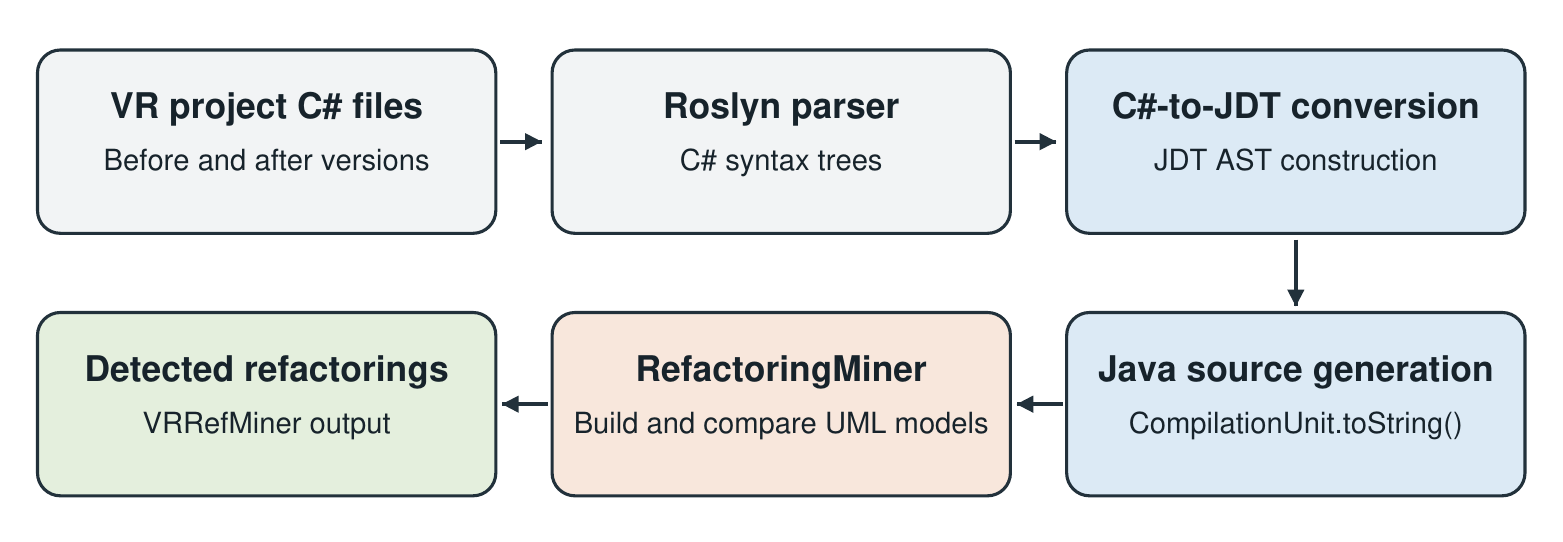}
    \caption{\tool{} Design Approach.}
    \label{fig:vrrefminer}
\end{figure}

To address these limitations, we develop \tool{}, which extends RefactoringMiner to support automated detection of conventional refactorings in C\# code from Unity-based VR projects. Our approach is inspired by Dilhara et al.~\cite{Dilhara2022}, who introduced JavaFyPy to enable the reuse of Java-based analysis tools for Python code through AST transformation. Following a similar approach, \tool{} transforms C\# syntax trees into a representation compatible with RefactoringMiner's existing analysis infrastructure. This enables refactoring detection in C\# without developing new matching algorithms. \autoref{fig:vrrefminer} presents the overall architecture of \tool{}, which we describe in detail below.

\subsubsection{\tool{} Design }
\label{subsubsec:pipeline}

The architecture of \tool{}, illustrated in~\autoref{fig:vrrefminer}, consists of two main components: (1) C\#-to-JDT AST transformation and (2) structural model construction and refactoring detection. We describe these components below, followed by an evaluation of the transformation's structural fidelity.

\noindent\textbf{1. C\#-to-JDT AST Transformation:}
\tool{} transforms C\# syntax trees produced by Microsoft Roslyn into Eclipse JDT ASTs using an intermediate representation and a transformation visitor. The transformation accommodates differences in declaration structures, expressions, and language-specific constructs while preserving structural information relevant to refactoring detection.

\begin{figure*}[t]
    \centering
    \includegraphics[width=\textwidth]
    {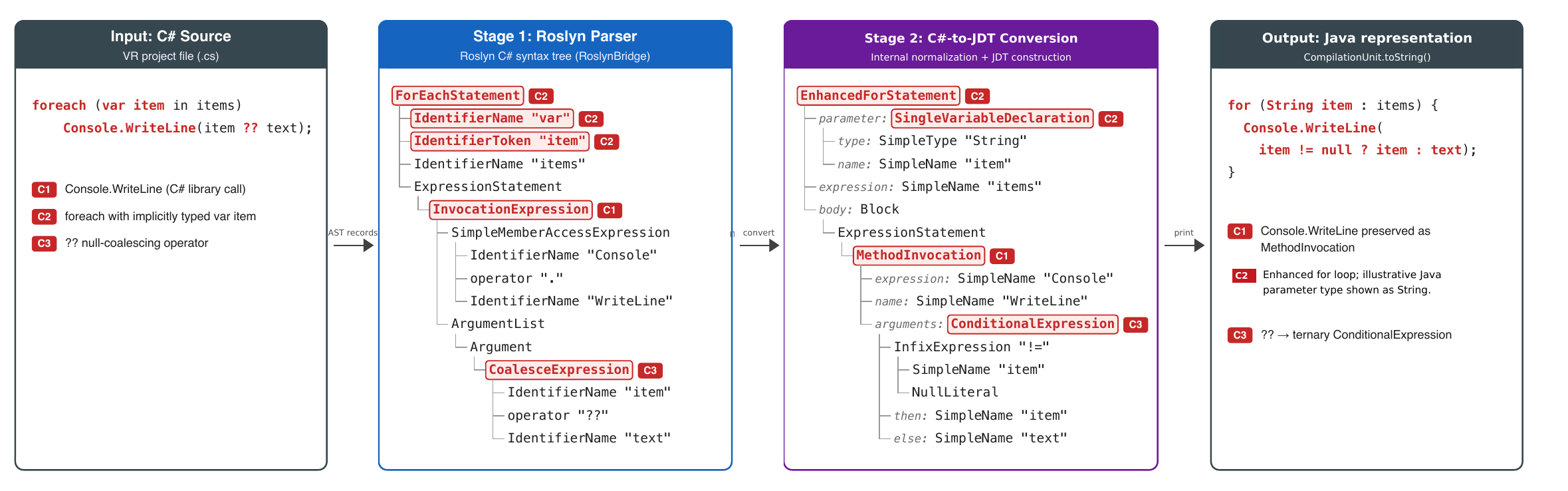}
    \caption{Example of Roslyn-based C\#-to-JDT AST transformation.
Simplified trees show the source representation, the internal
conversion representation, and the resulting JDT AST.
Labels C1--C3 track selected constructs across the conversion.}
    \vspace{-0.5cm}
    \label{fig:pipeline}
\end{figure*}

\paragraph{Tree Representation:} We model the source and target ASTs as rooted, ordered trees $\mathcal{T}=(\mathcal{V},\mathcal{E})$, where each node $v\in\mathcal{V}$ has a syntactic kind $\tau(v)$, an optional label $\ell(v)$, and an ordered sequence of children. Let $\mathcal{T}^{R}$ and $\mathcal{T}^{J}$ denote the Roslyn-based source tree and the resulting Eclipse JDT tree, respectively. We define the transformation as:
\[
F : \mathcal{T}^{R} \longrightarrow \mathcal{T}^{J}.
\]

The transformation consists of two steps: (1) normalizing source nodes into an intermediate representation compatible with the transformation visitor and (2) constructing corresponding JDT nodes for declarations, statements, and expressions. This process operates on the parsed C\# representation without reparsing the original source code.

\paragraph{Node Correspondence:} Because C\# and Java differ in their syntactic structures, the transformation does not necessarily establish a one-to-one mapping between source and target nodes. We represent their structural correspondence using:
\[
\Phi\subseteq\mathcal{V}^{R}\times\mathcal{V}^{J}.
\]
A source construct may map to multiple target nodes, while a target construct may incorporate information from multiple source nodes. We distinguish three mapping strategies:

\begin{itemize}
\item \textbf{Direct correspondence:}
Constructs with compatible representations retain their relevant syntactic information. For example, the invocation labeled C1 in~\autoref{fig:pipeline} is mapped to a JDT \texttt{MethodInvocation}, preserving its receiver, method name, and argument structure.

\item \textbf{Structural adaptation:}
Constructs with different syntactic organizations are transformed into corresponding JDT structures. For example, the C\# \texttt{foreach} statement (C2) is mapped to an \texttt{EnhancedForStatement}, preserving its iteration variable, collection expression, and body. Similarly, the null-coalescing expression (C3) is represented using a \texttt{ConditionalExpression} with an explicit null check. This mapping applies to the simple variable operand illustrated in the figure.

\item \textbf{Language-specific approximation:}
C\# constructs without direct Java equivalents are represented using supported JDT structures. For example, properties are mapped to fields and accessor methods, while structs are represented as classes containing their members. These approximations retain structural information needed for analysis but do not preserve all language-specific semantics, such as C\# value-type behavior.

\end{itemize}

~\autoref{fig:pipeline} illustrates the transformation of a C\# \texttt{foreach} statement containing a null-coalescing expression into an Eclipse JDT AST. The figure highlights how the intermediate representation supports the mapping of C\# constructs to corresponding or adapted JDT structures. The resulting \texttt{CompilationUnit} provides a Java-oriented representation suitable for RefactoringMiner's structural analysis rather than an executable translation of the original C\# program.

\noindent\textbf{2. Structural Model Construction and Refactoring Detection:}
Following AST transformation, \tool{} serializes each JDT \texttt{CompilationUnit} into Java-oriented source text using \texttt{cu.toString()}. The generated text is passed to RefactoringMiner's existing \texttt{UMLModelASTReader}, which reparses it using Eclipse JDT and constructs structural models for the before and after revisions. This step enables the transformed C\# representation to be processed through RefactoringMiner's existing analysis pipeline. RefactoringMiner subsequently compares these models using its existing model differencing and refactoring detection algorithms~\cite{tsantalis2018accurate}. Thus, \tool{} extends the input processing pipeline to support C\# while reusing the underlying detection algorithms. Reported source locations correspond to the generated Java-oriented representation; mapping them to the original C\# source files requires additional source-location mapping.

\noindent\textbf{3. Transformation Validation:}
To assess the structural correctness of the C\#-to-JDT AST transformation, we designed a construct-based validation using 30 hand-written C\# fixtures. The fixtures cover eight groups of language constructs: declarations, class members, method signatures, control flow, expressions, C\#-specific constructs, type relationships, and Unity-related constructs. For each fixture, we inspected the resulting JDT representation against the expected construct, including its relevant names, operands, and structural relationships. A test was considered fully passed when its required elements were preserved, partially passed when only some information was retained or an approximation was introduced, and failed when the required structure was missing. Of the 30 fixtures, 24 fully passed and six were partial; none failed within this core test set. These results demonstrate correct structural mappings for the 24 fully passed fixtures, but do not establish complete semantic equivalence for the six approximations.

To assess whether the transformed representations could support refactoring detection, we constructed six controlled before-and-after scenarios covering Rename Method, Extract Method, Move Method, Rename Class, Extract Class, and Move Class. \tool{} detected the expected refactoring type in all six scenarios, although two also produced additional detections. Five non-refactoring controls produced no unexpected detections. Four separate failure-oriented tests exposed one failed preservation case, one unsupported construct, and two partial transformations. These results provide controlled evidence of structural correctness for selected constructs and functional detection of selected conventional refactorings, while identifying limitations that prevent claims of complete transformation correctness or general detection accuracy. The detailed fixtures and results are provided in the replication package.

\section{Empirical Evaluation}
\label{sec:evaluation}
This section investigates refactoring types (RQ1), their relationship to technical debt (RQ2), and the effectiveness of \tool{} in detecting conventional and VR-specific refactorings (RQ3).


\subsection{{RQ1: Types of Refactoring in VR Projects}}
\label{subsec:rq1_results}
\begin{figure*}[t]
    \centering
    \includegraphics[width=0.9\linewidth,height=6.5cm]{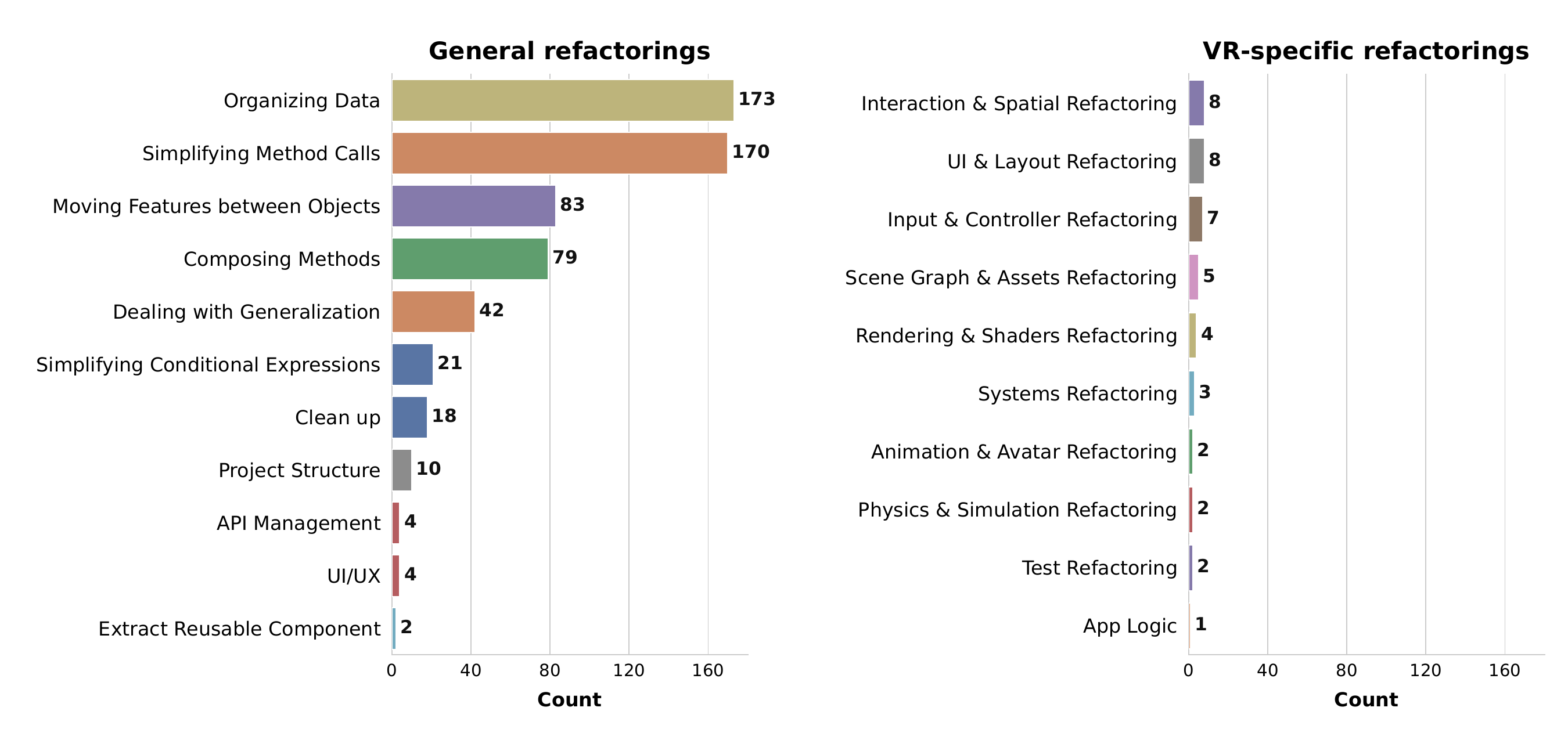}
    \caption{Distribution of Refactoring Types in VR Projects.}
    \vspace{-0.3cm}
    \label{fig:RQ1_Sunburst}
\end{figure*}

To answer \textbf{RQ1}, we analyzed 303 commits selected from VR software projects using a sample-size-based \emph{random sampling} procedure with a 95\% confidence level and a 5\% margin of error, as detailed in~\autoref{subsec:manual}. Based on our manual analysis, we identified two categories of refactoring types: (i) general refactorings and (ii) VR-specific refactorings. General refactorings, such as Organizing Data and Simplifying Method Calls, are common across software systems and also occur in VR projects. Importantly, applying a general refactoring to a VR-related subsystem does not, by itself, constitute a VR-specific refactoring. In contrast, VR-specific refactorings, such as Input \& Controller Refactoring, address domain-specific requirements. The overall distribution of both categories is shown in~\autoref{fig:RQ1_Sunburst}.

Across the \textbf{303} sampled commits, we identified \textbf{648} refactoring instances, as individual commits may contain multiple refactoring operations. Of these, \textbf{33} commits contained no refactorings, leaving \textbf{270} commits with at least one refactoring. Among the identified instances, \textbf{606} (\textbf{93.52\%}) were general refactorings, while \textbf{42} (\textbf{6.48\%}) were VR-specific. These findings indicate that general refactorings dominate VR software maintenance, although VR-specific refactorings also occur.

\subsubsection{General Software Engineering Refactorings.}

Among the \textbf{606} general refactoring instances, \textit{Organizing Data} was the most frequent category (\textbf{173}), followed by \textit{Simplifying Method Calls} (\textbf{170}) and \textit{Moving Features between Objects} (\textbf{83}). The first two categories collectively accounted for \textbf{56.60\%} of general refactorings, indicating that data organization and method-level modifications are prominent maintenance activities in VR projects. Within \textit{Simplifying Method Calls}, \textit{Rename Method} (52) and \textit{Change Method Access Modifier} (35) were frequently observed. Similarly, \textit{Moving Features between Objects} included \textit{Extract Class} (22) and \textit{Move Attribute} (13), reflecting changes to class responsibilities and the placement of data.




\autoref{lst:move-method} illustrates a \textit{Move Method} refactoring in a Unity-based VR project. In this example, configuration toggle-handling functionality was moved from individual editor windows into a generic base class, \texttt{AutoConfigureWindow<TSetting>}. This change centralizes shared UI logic, reduces duplication across configuration windows, and improves the reusability of editor functionality. Although the refactoring occurs in a VR project, the underlying operation is a conventional \textit{Move Method} refactoring rather than a VR-specific refactoring.


\begin{lstlisting}[
  language=Diff,
  basicstyle=\scriptsize\ttfamily,
  frame=single,
  caption={Move Method refactoring to centralize configuration toggle logic in a generic VR editor window from \texttt{RussellXie7/Unity\_Hololens\_Dev.f}},
  label={lst:move-method}
]
- public class AutoConfigureWindow : UnityEditor.EditorWindow
+ public abstract class AutoConfigureWindow<TSetting> : UnityEditor.EditorWindow
  {
+   private Dictionary<TSetting, bool> values = new Dictionary<TSetting, bool>();
+   ...
+   }
    protected virtual void OnGUI()
    {
+     var keys = values.Keys.ToArray();
+     ...
  }
\end{lstlisting}

Less frequent general refactorings included \textit{Inline/Temporary Variable Refactoring} (12) and \textit{Replace Variable With Attribute} (8). Together, these findings suggest that general refactoring activities in VR projects primarily involve reorganizing data, modifying method interfaces, and redistributing responsibilities across classes. We further examine their relationship to technical debt categories in RQ2.

\subsubsection{VR-specific Refactorings}
VR-specific refactorings (\textbf{n=42}) account for \textbf{6.48\%} of all observed refactoring instances. The most frequent categories are \emph{Interaction \& Spatial Refactoring} (8) and \emph{UI \& Layout Refactoring} (8), followed by \emph{Input \& Controller Refactoring} (7), \emph{Scene Graph \& Assets Refactoring} (5), and \emph{Rendering \& Shaders Refactoring} (4). Less frequent categories include \emph{Systems Refactoring} (3), \emph{Animation \& Avatar Refactoring} (2), \emph{Physics \& Simulation Refactoring} (2), \emph{Test Refactoring} (2), and \emph{App Logic} (1). This distribution highlights the prominence of immersive interaction, spatial interfaces, controller input, and rendering-related concerns in VR-specific refactoring activities. Table~\ref{tab:vr_refactoring_definitions}
summarizes the ten VR-specific refactoring
categories identified through our manual analysis
and provides their corresponding definitions.

\begin{table*}[t]
\centering
\footnotesize
\caption{Definitions of VR-Specific Refactoring Categories}
\begin{tabularx}{\textwidth}{@{}p{0.27\textwidth}X@{}}
\hline
\textbf{VR-Specific Category} & \textbf{Definition} \\
\hline

Interaction \& Spatial Refactoring &
Restructuring how VR objects and components respond to user actions or manage spatial relationships, including object selection, manipulation, ray-based interaction, and spatial positioning. \\

UI \& Layout Refactoring &
Reorganizing VR user-interface components, spatial UI layouts, menus, or presentation structures to improve modularity and maintainability. \\

Input \& Controller Refactoring &
Restructuring how VR systems acquire, map, process, or dispatch input from controllers, tracking devices, and other input sources while preserving the intended input semantics. \\

Scene Graph \& Assets Refactoring &
Reorganizing VR scene hierarchies, prefab structures, serialized assets, or component relationships while preserving intended scene functionality. \\

Rendering \& Shaders Refactoring &
Restructuring VR rendering components, shader organization, camera handling, or clipping logic to improve maintainability and organization. \\

Systems Refactoring &
Reorganizing VR runtime systems, supporting utilities, or subsystem dependencies to improve modularity, consistency, or maintainability. \\

Animation \& Avatar Refactoring &
Restructuring avatar-related components, animation handling, interpolation utilities, or motion-related logic while maintaining intended behavior. \\

Physics \& Simulation Refactoring &
Reorganizing VR physics, collision handling, simulation components, or object-interaction logic while preserving intended simulation behavior. \\

Test Refactoring &
Restructuring tests, test utilities, or testing-related resources associated with VR components to improve maintainability and test organization. \\

App Logic Refactoring &
Restructuring VR application-level logic or supporting components that coordinate application functionality without introducing new intended behavior. \\

\hline
\end{tabularx}

\vspace{-0.6cm}
\label{tab:vr_refactoring_definitions}
\end{table*}



Unlike general refactorings, VR-specific refactorings address maintenance concerns associated with immersive interactions, spatial interfaces, and runtime behavior. For example, \textit{Interaction \& Spatial Refactoring} concerns 3D object manipulation, ray-based interaction, and spatial transformations. \textit{UI \& Layout Refactoring} involves reorganizing immersive interfaces, including world-space UI elements and viewpoint-dependent layouts. \textit{Input \& Controller Refactoring} addresses VR controller mappings, motion-tracked input, trigger and grip interactions, haptic feedback, and input-device abstractions. Similarly, \textit{Scene Graph \& Assets Refactoring} involves Unity GameObject hierarchies, prefab organization, transform relationships, and serialized scene configurations. \textit{Rendering \& Shaders Refactoring} concerns camera configurations, projection modes, shader organization, and runtime rendering updates. These categories capture VR-oriented maintenance concerns that extend beyond conventional source-code restructuring, particularly when changes involve relationships between C\# code, scene assets, and runtime behavior.



\begin{lstlisting}[language=Diff, basicstyle=\scriptsize\ttfamily, frame=single,
caption={Scene- and prefab-level refactoring using tag-based activation to externalize runtime behavior into Unity configuration artifacts from \texttt{takafumihoriuchi/SmartCodesignVR.}},
label={lst:tag-based-activation}]
Assets/myScenes/PrototypingScene.unity
-  m_TagString: Untagged
+  m_TagString: ActivateOnLoad
ProjectSettings/TagManager.asset
   tags:
   - DeactivateOnLoad
+  - ActivateOnLoad
/Assets/myScripts/ScriptsPrototypingScene/PrototypingSceneCore.cs
+    private void ActivateTaggedObjects()
+    {
+        var taggedObjects = GameObject.FindGameObjectsWithTag("ActivateOnLoad");
+        foreach (var obj in taggedObjects) obj.SetActive(true);
+    }
\end{lstlisting}

Listing~\ref{lst:tag-based-activation} illustrates a \textit{Scene Graph \& Assets Refactoring} example from the SmartCodesignVR project. In this commit, developers replaced static object activation settings with a tag-driven runtime mechanism. Specifically, objects were assigned an \texttt{ActivateOnLoad} tag in Unity's serialized scene configuration, while a new \texttt{ActivateTaggedObjects()} method dynamically activates tagged objects at runtime. This restructuring centralizes object activation logic and enables scene metadata to control runtime behavior. The example highlights how VR-specific refactorings can span both C\# source code and Unity scene artifacts, requiring coordinated changes to scene configuration and runtime logic.
\begin{figure}[t]
    \centering
    \includegraphics[width=\columnwidth]{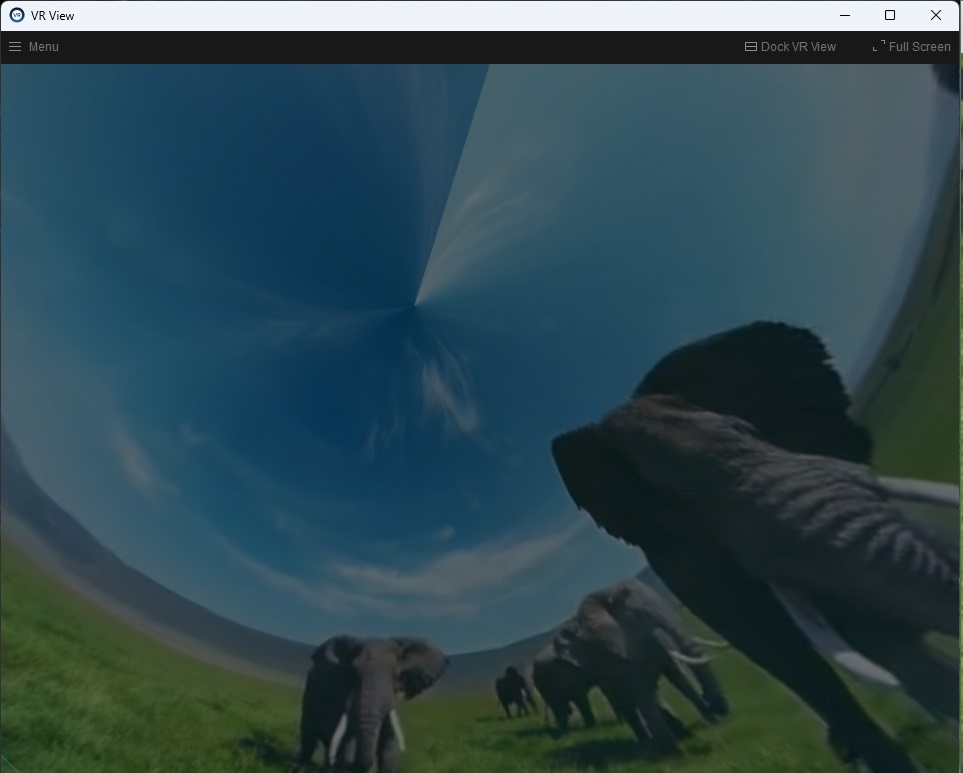}
    \caption{illustrates an example of VR software engineering refactoring in the BIVROST/360PlayerWindows project}
    \label{fig:vr_visual}
\end{figure}

~\autoref{fig:vr_visual} illustrates a \textit{Projection \& Display Mode Management} refactoring in commit \textit{150f71}~\cite{displayrefactor} from the \textit{BIVROST/360PlayerWindows} project. The refactoring consolidates projection handling within \texttt{MediaDecoder}, replacing fragmented updates across UI and rendering components with a centralized mechanism. It eliminates redundant update paths and per-frame polling, introduces an explicit \texttt{Autodetect} mode, and updates rendering geometry only when the projection mode or content changes. These changes improve coordination between media decoding and rendering components, illustrating how VR-specific refactoring addresses runtime projection switching, display consistency, and rendering-related maintenance concerns.



Overall, VR-specific refactorings reveal maintenance concerns associated with spatial interactions, motion-tracked input, immersive interfaces, scene organization, and real-time rendering. Although they account for a smaller proportion of observed refactorings, these changes highlight the importance of cross-artifact relationships and runtime context in VR software maintenance. Such characteristics also expose limitations of conventional code-centric refactoring detection approaches, which primarily analyze structural changes in source code.

\subsection{RQ2: Types of Technical Debt Addressed by Refactoring}
\label{subsec:rq2_results}
\vspace{-0.1cm}
RQ2 investigates the relationship between the refactoring types identified in RQ1 and technical debt in VR software projects. Using the manually validated refactoring data described in~\autoref{subsec:manual}, we follow prior domain-specific research on refactoring and technical debt in machine learning systems~\cite{tang2021mltdrefactoring} to characterize technical debt associated with refactoring activities in VR software.


\subsubsection{Overview of Technical Debt Categories}
\begin{figure}[t]
    \centering
    \includegraphics[width=\linewidth]{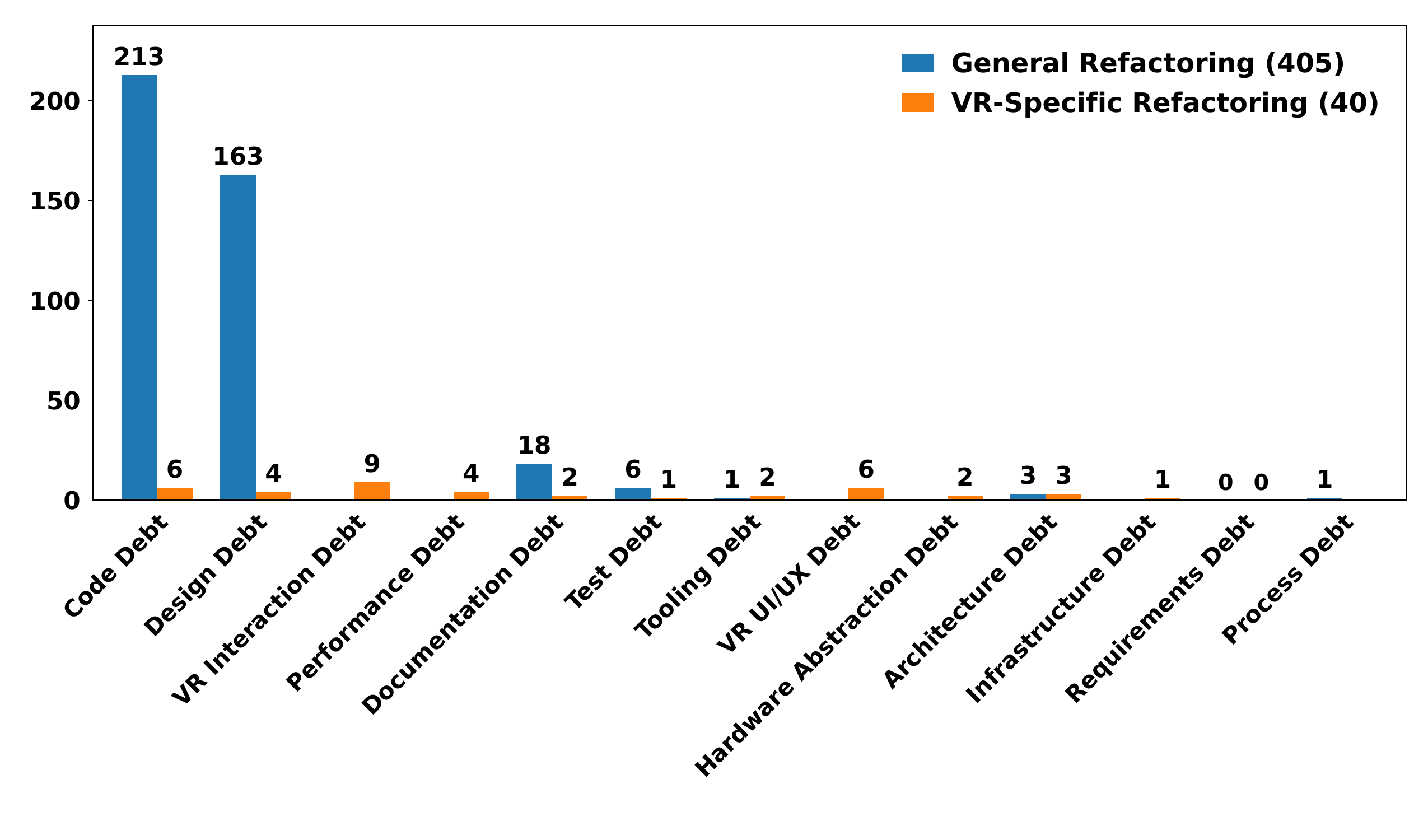}
     \vspace{-0.7cm}
    \caption{Distribution of Technical Debt Categories.}
     \vspace{-0.1cm}
    \label{fig:rq2_tech_debt}
\end{figure}

Figure~\ref{fig:rq2_tech_debt} presents the distribution of technical debt categories associated with general and VR-specific refactorings. Overall, \textit{Code Debt} is the most frequently observed category, accounting for \textbf{219} of the \textbf{445} debt-label assignments. We employed a multi-label classification scheme, allowing each refactoring commit to be associated with multiple technical debt categories.

\begin{table*}[t]
\centering
\footnotesize
\caption{Definitions of VR-Specific Technical Debt Categories}
\begin{tabularx}{\textwidth}{@{}p{0.27\textwidth}X@{}}
\hline
\textbf{VR-Specific Technical Debt Category} &
\textbf{Definition} \\
\hline

VR Interaction Debt &
Technical debt arising from poorly structured or
inconsistent VR interaction mechanisms, including
spatial object manipulation, ray-based interactions,
haptic feedback, interaction-state management,
and input-response timing. \\

VR UI/UX Debt &
Technical debt arising from inconsistencies in
immersive user interfaces and experiences, including
spatial UI layouts, visual transitions, feedback
consistency, and spatial coherence. \\

Hardware Abstraction Debt &
Technical debt arising from fragmented, inconsistent,
or tightly coupled abstractions for VR hardware,
including controller mappings, tracking-device
interfaces, device-specific dependencies, and
hardware API integration. \\

VR Infrastructure Debt &
Technical debt arising from inadequate or
difficult-to-maintain VR development environments,
device configurations, SDK integrations, calibration
setups, and hardware-dependent development or
testing infrastructure. \\

\hline
\end{tabularx}

\vspace{-0.4cm}
\label{tab:vr_td_definitions}
\end{table*}


Table~\ref{tab:vr_td_definitions}
presents the definitions of the proposed
VR-specific technical debt categories
identified through our manual analysis. We identified \textbf{445} debt-label assignments across \textbf{270} commits containing refactorings, including \textbf{405} assignments associated with general refactorings and \textbf{40} with VR-specific refactorings. To examine relationships between refactoring types and technical debt categories, we further analyzed their co-occurrences, yielding \textbf{761} technique--debt links for general refactorings and \textbf{59} for VR-specific refactorings. As shown in Table~\ref{tab:tech_debt}, notable associations include \textit{UI \& UX Refactoring} with \textit{Code Debt} (3/3 links; OR = 6.43) and \textit{Dealing with Generalization} with \textit{Tooling Debt} (1/63 links; OR = 5.61). Among VR-specific refactorings, associations include \textit{Test Refactoring} with \textit{Test Debt} (1/1 link; OR = 67.80), \textit{Scene Graph \& Assets Refactoring} with \textit{Code Debt} (4/7 links; OR = 33.33), and \textit{Animation \& Avatar Refactoring} with \textit{Test Debt} (1/3 links; OR = 13.50). Although some associations yield large odds ratios, several are based on sparse observations and should therefore be interpreted as exploratory rather than robust statistical relationships.

\begin{table}[t]
\centering
\scriptsize
\caption{Top technique–debt associations by odds ratio, with zero-cell correction where needed.}
\label{tab:tech_debt}

\renewcommand{\arraystretch}{1.12}
\setlength{\tabcolsep}{3pt}


\begin{tabular}{l l c c r}
\toprule
\shortstack{\textbf{Technique}\\\textbf{category}} &
\shortstack{\textbf{Debt}\\\textbf{category}} &
\shortstack{\textbf{Within}\\\textbf{technique}}&
\shortstack{\textbf{Within} \\\textbf{debt}} &
\textbf{OR} \\
\midrule

\multicolumn{5}{l}{%
\textbf{General refactorings (Total links: 761)}} \\
\midrule

UI/UX & Code Debt & 3/3 & 3/398 & 6.43 \\

Dealing with Generalization & Tooling Debt & 1/63 & 1/3 & 5.61 \\

Composing Methods & Process Debt & 1/120 & 1/2 & 5.38 \\

Moving Features between Objects & Tooling Debt & 1/103 & 1/3 & 3.22 \\

Organizing Data & Process Debt & 1/210 & 1/2 & 2.63 \\

\midrule
\multicolumn{5}{l}{%
\textbf{VR-specific refactorings (Total links: 59)}} \\
\midrule

Test Refactoring & Test Debt & 1/1 & 1/3 & 67.80 \\

Scene Graph \& Assets Refactoring & Code Debt & 4/7 & 4/6 & 33.33 \\

App Logic & Documentation Debt & 1/3 & 1/2 & 27.50 \\

Animation \& Avatar Refactoring & Test Debt & 1/3 & 1/3 & 13.50 \\

Systems Refactoring & Tooling Debt & 2/7 & 2/4 & 10.00 \\

Physics \& Simulation Refactoring & VR Interaction Debt & 1/1 & 1/15 & 9.21 \\

\bottomrule
\end{tabular}%
\end{table}

\subsubsection{General Refactoring Tech Debt Patterns}
Among the 405 debt-label assignments associated with general refactorings, \textit{Code Debt} (213) and \textit{Design Debt} (163) collectively account for 376 assignments (92.84\%). This concentration indicates that general refactorings in VR projects primarily involve code organization and software design concerns, reflecting the maintenance of evolving Unity-based codebases. Other categories, including \textit{Documentation Debt} (18), \textit{Test Debt} (6), \textit{Architecture Debt} (3), \textit{Tooling Debt} (1), and \textit{Process Debt} (1), occur less frequently.

The co-occurrence analysis further reveals how general refactoring families relate to different technical debt categories. For example, \textit{UI/UX Refactoring} co-occurs with \textit{Code Debt} (3/3 links; OR = 6.43), illustrating that UI-related maintenance can involve underlying code organization rather than visual presentation alone. \textit{Tooling Debt} co-occurs with \textit{Dealing with Generalization} (1/63; OR = 5.61) and \textit{Moving Features between Objects} (1/103; OR = 3.22), while \textit{Process Debt} appears alongside \textit{Composing Methods} (1/120; OR = 5.38) and \textit{Organizing Data} (1/210; OR = 2.63). Although these associations suggest that conventional refactorings can accompany broader development concerns, the tooling- and process-related findings each rely on a single link and should be interpreted cautiously. Overall, the findings indicate that general refactorings in VR projects are predominantly associated with code and design debt, with less frequent associations involving tooling and process concerns.



\subsubsection{VR-Specific Technical Debt Patterns}

Beyond general technical debt categories, our analysis identifies VR-specific concerns related to user interactions, interface design, and runtime behavior. Among these, \textit{VR Interaction Debt} (9) and \textit{VR UI/UX Debt} (6) are the most frequently observed VR-specific debt categories, as shown in Figure~\ref{fig:rq2_tech_debt}. Although the co-occurrence analysis identifies 15 links involving \textit{VR Interaction Debt}, only one is associated with \textit{Physics \& Simulation Refactoring} in the reported link view. This indicates that interaction-related debt spans multiple refactoring categories rather than being concentrated in a single category.


\begin{lstlisting}[
    language=diff,
    basicstyle=\scriptsize\ttfamily,
    frame=single,
    framesep=4pt,
    caption={VR interaction debt caused by numerically unstable random input generation from \texttt{gpvigano/VRTK-GearVR-Test.}},
    label={lst:vr-interaction-debt-random}
]
/Assets/NormalizedRandom.cs
-  float x = Random.Range(0.0f, 1.0f);
-  float y = Random.Range(0.0f, 1.0f);
+  float x = Random.Range(0.01f, 0.99f);
+  float y = Random.Range(0.01f, 0.99f);
\end{lstlisting}
\vspace{-0.3cm}

~\autoref{lst:vr-interaction-debt-random} illustrates a VR interaction-related change from the \texttt{gpvigano/VRTK-GearVR-Test} project. The developers modified calls to \texttt{Random.Range()} by restricting the generated values from the interval [0.0, 1.0] to [0.01, 0.99]. This change excludes values near the interval boundaries, potentially avoiding edge cases in subsequent interaction-related computations. The example illustrates how VR-specific maintenance can involve numerical adjustments to interaction-related logic. In contrast to interaction and UI/UX debt, performance-related debt appears less frequently in our analyzed refactoring commits.

\subsection{RQ3: Effectiveness of VRRefMiner Detection}
\label{subsec:rq3_results}
\vspace{-0.1cm}
We evaluate the effectiveness of \tool{} in detecting refactorings in Unity-based VR projects using the manually labeled dataset (~\autoref{subsec:manual}) and the full mined dataset (~\autoref{subsec:data}). Specifically, we assess its ability to detect conventional refactorings in C\# code and identify its limitations in recognizing VR-specific refactoring categories.


\subsubsection{Effectiveness of \toolminer{} on Labeled Data}
\vspace{-0.1cm}
We evaluated \tool{} using the manually labeled dataset comprising 270 refactoring commits, including 246 commits containing general refactorings and 24 containing VR-specific refactorings. As shown in Table~\ref{tab:rq3_sample_summary}, \tool{} detected at least one refactoring in 216 of the 246 general-refactoring commits (87.80\%), leaving 30 commits without detections.

To assess detection accuracy at the refactoring-type level, we compared \tool{}'s results against the manually established ground truth. We evaluated all 246 general-refactoring commits
by matching distinct manually labeled refactoring
types against tool detections within each commit.
Repeated occurrences of the same type were counted
once per commit, unlike the instance-level counts
in RQ1. This yielded 185 TP, 36 FP, and 61 FN. Corresponding to a Precision of 83.71\%, recall of 75.20\%, F1  of 79.23\%. These metrics characterize the tool's effectiveness in detecting conventional structural refactorings in C\# code. For the 24 commits containing VR-specific refactorings, \tool{} did not identify any corresponding VR-specific refactoring category (0/24). This limitation reflects the tool's reliance on conventional structural refactoring detection, which does not explicitly recognize domain-specific transformations involving VR interaction logic, Unity components, scene configurations, and cross-artifact relationships~\cite{negara2013mining,rzig2023vrtesting,bierbaum2003vrtesting}.

\begin{table}[t]
\centering
\scriptsize
\caption{\tool{} Performance on Manually Labeled Data.}
\label{tab:rq3_sample_summary}
\begin{tabular}{lrr}
\toprule
\textbf{Category} & \textbf{\# Commits} & \textbf{Tool Detected} \\
\midrule
Refactoring commits (final analyzed set) & 270 & -- \\
\quad General SE refactoring & 246 & 216 (87.80\%) \\
\quad VR-specific refactoring & 24 & 0 (Domain/Tool Gap) \\
\bottomrule
\end{tabular}
\end{table}

\subsubsection{\tool{} Detection Results on the Full Dataset}
We applied \tool{} to the full dataset of 1,755 potential refactoring commits collected from Unity-based VR projects. The tool detected at least one refactoring in 1,281 commits (72.99\%), identifying 15,549 refactoring operations across 68 distinct types. Among commits with detections, the average number of detected refactorings was 12.14, with a median of 6 and a maximum of 380. Table~\ref{tab:detected_refactoring_types} summarizes the distribution of detected refactoring types. Structural moves and extractions constituted the largest category (31.0\%), followed by renaming operations (19.7\%), access modifier changes (18.7\%), type changes (15.6\%), and parameter operations (9.3\%). These findings indicate that \tool{} identifies a diverse range of conventional refactoring operations in Unity-based VR codebases, particularly those involving code reorganization, naming changes, and modifications to method and type declarations. Overall, the results demonstrate the applicability of \tool{} to conventional structural refactoring detection in C\# code. However, the detected operations are limited to refactoring types supported by the underlying structural analysis and do not explicitly capture VR-specific transformations involving interaction logic, scene configurations, or cross-artifact relationships. This distinction highlights the need for domain-aware refactoring detection techniques tailored to VR software.

\begin{table}[t]
\centering
\footnotesize
\caption{Categorization of 15,549 refactorings detected by \tool{} across 1,281 VR/Unity commits (68 distinct types).}
\label{tab:detected_refactoring_types}
\begin{tabular}{p{1.8cm} r r p{4.0cm}}
\toprule
\textbf{Category} & \textbf{Count} & \textbf{\%} & \textbf{Top Detected Types} \\
\midrule
Structural moves/extracts & 4,819 & 31.0\% & Move Attribute (902), Move Method (996), Extract Method (515), Extract Class (209), Extract And Move Method (326) \\
Renaming & 3,061 & 19.7\% & Rename Method (1,141), Rename Attribute (798), Rename Variable (569), Rename Parameter (370), Rename Class (176) \\
Access modifiers & 2,904 & 18.7\% & Change Method Access Modifier (1,043), Change Attribute Access Modifier (916), Add Method Modifier (158), Add Attribute Modifier (394) \\
Type changes & 2,426 & 15.6\% & Change Variable Type (702), Change Attribute Type (702), Change Parameter Type (588), Change Return Type (430) \\
Parameter operations & 1,441 & 9.3\% & Add Parameter (734), Remove Parameter (489), Reorder Parameter (17) \\
Variable/attribute ops & 476 & 3.1\% & Parameterize Variable (105), Replace Variable With Attribute (62), Replace Attribute With Variable (66), Encapsulate Attribute (161) \\
Inline operations & 238 & 1.5\% & Inline Variable (144), Inline Method (66), Inline Attribute (7) \\
Conditional operations & 96 & 0.6\% & Invert Condition (47), Merge Conditional (28), Split Conditional (14), Replace Conditional With Ternary (7) \\
Merge/Split operations & 85 & 0.5\% & Merge Attribute (13), Merge Class (8), Split Class (4), Merge Parameter (8), Split Method (2) \\
Other & 3 & 0.0\% & Move Code, Add Method Annotation \\
\midrule
\textbf{Total} & \textbf{15,549} & \textbf{100\%} & \textbf{68 distinct refactoring types} \\
\bottomrule
\end{tabular}
\end{table}

\noindent\textbf{Coverage Limitations and Detection Gaps.}
Despite its effectiveness in detecting conventional structural refactorings, \tool{} has three key limitations in VR/Unity projects. First, \textit{Unity-specific abstractions}, such as \texttt{MonoBehaviour} lifecycle methods, coroutines, and scene management, are not explicitly modeled, causing domain-specific transformations to be classified as conventional structural changes or remain undetected. Second, \textit{cross-artifact refactorings} involving C\# scripts, prefabs, scenes, and configuration files cannot be fully captured through source-code analysis alone. Third, \textit{semantic transformations}, such as restructuring interaction state machines or event-handling logic, may not correspond to recognizable refactoring patterns. These limitations highlight the need for domain-aware detection techniques that integrate source-code analysis with Unity-specific artifacts.

\vspace{-0.2cm}
\section{Threats to Validity}
\label{sec:threats}
\vspace{-0.2cm}
\noindent\textbf{Construct Validity:}
Manual labeling may introduce human error and subjectivity, particularly in identifying VR-specific refactorings and their associations with technical debt. We classified refactorings based on the nature of changes rather than repository context. To mitigate potential misclassification, multiple researchers experienced in VR development and research independently reviewed the labels and resolved disagreements through consensus

\noindent\textbf{Internal Validity:} While \tool{} assists in detecting refactorings, it may not capture all semantic, cross-artifact, or Unity/VR-specific changes. As a result, some refactorings could be missed, potentially affecting the completeness of the observed patterns and insights. To mitigate this limitation, we complemented the automated detection with thorough manual analysis and observation.

\noindent\textbf{External Validity:} This study focuses exclusively on Unity-based VR projects, limiting the generalizability of results to other game engines, VR platforms, or interactive software domains. The analysis is based on 314 open-source Unity projects with more than 100 commits, many of which are backed by large organizations. As a result, the dataset may not capture all coding styles, project sizes, or types of VR applications, though it reflects active and well-maintained Unity-based VR development. Furthermore, by filtering out the commits with refactor and excluding commits that modify more than 20 source files may omit larger restructuring activities. Consequently, the observed distributions may not fully represent all refactoring practices in VR software projects, particularly those involving large-scale or cross-artifact changes.

\vspace{-0.1cm}
\section{Implications}
\label{sec:implications}
\vspace{-0.1cm}
Our analysis of refactorings and technical debt in Unity-based VR projects yields several implications for developers, tool builders, and researchers.

\noindent\textbf{VR Developers:} General and VR-specific refactorings highlight that Code and Design Debt, as well as Interaction and UI/UX Debt, are prevalent in VR projects. Developers should prioritize addressing these debts by stabilizing interaction logic, reducing redundant or tightly coupled code, and improving maintainability. Awareness of these patterns can guide better design decisions and more effective refactoring practices in VR systems.

\noindent\textbf{VR Tool Builders:} Our findings reveal that existing refactoring tools often fail to capture VR-specific changes, especially those involving cross-artifact or engine-dependent logic. Tool developers should consider extending automated refactoring detection to handle VR-specific patterns, real-time interaction pipelines, and performance-critical code, enabling more comprehensive support for VR maintenance tasks.

\noindent\textbf{Software Engineering Researchers:} The study demonstrates that VR software presents unique refactoring and technical debt challenges distinct from traditional systems. Researchers can investigate domain-specific refactoring techniques, automated detection approaches, and methods to mitigate VR-specific debts. The observed patterns can inform studies on maintainability, code quality, and engineering practices in interactive 3D systems. Additionally, researchers should develop VR-aware maintainability metrics (e.g., scene complexity, prefab coupling, script–asset dependency density) and examine how they relate to refactoring frequency, defects, and performance regressions.

\vspace{-0.2cm}
\section{Conclusion}
\label{sec:conclusion}
\vspace{-0.1cm}
In this paper, we conducted the first empirical study of refactoring practices and technical debt in Unity-based VR projects, analyzing 314 open-source projects through both manual analysis and automated detection (\tool{}). Our results show that general refactorings primarily address Code and Design Debt, while VR-specific refactorings focus on Interaction and UI/UX Debt, reflecting the unique challenges of real-time 3D interactions and engine-level coupling. We also identified patterns in how refactorings mitigate unstable interaction logic and redundant code, and improve maintainability, providing actionable insights for VR development. However, current refactoring tools and algorithms remain insufficient to detect VR-domain-specific changes.

These findings emphasize the need to treat VR systems as a distinct domain for refactoring and technical debt analysis, where traditional software engineering practices must be adapted to real-time interactions, performance constraints, and cross-artifact dependencies. By combining automated and manual approaches, our study lays a foundation for improved refactoring detection, maintenance strategies, and quality assurance in VR software. Future work can extend this analysis to other VR engines and frameworks and explore automated techniques for mitigating VR-specific technical debt.

\bibliographystyle{IEEEtran}
\bibliography{biblio}

@String{Computing = "Computing" }

@inproceedings{tsantalis2018accurate,
  author    = {Nikolaos Tsantalis and Matin Mansouri and Laleh M. Eshkevari and Davood Mazinanian and Danny Dig},
  title     = {Accurate and Efficient Refactoring Detection in Commit History},
  booktitle = {Proceedings of the 40th International Conference on Software Engineering (ICSE)},
  year      = {2018}
}

@article{tsantalis2020refactoringminer,
  author  = {Nikolaos Tsantalis and Ameya Ketkar and Danny Dig},
  title   = {RefactoringMiner 2.0: A Multi-language Refactoring Detection Tool},
  journal = {IEEE Transactions on Software Engineering},
  year    = {2020},
  doi     = {10.1109/TSE.2020.3007721}
}

@inproceedings{tang2021mltdrefactoring,
  author    = {Yiming Tang and Raffi Khatchadourian and Mehdi Bagherzadeh and Rhia Singh and Ajani Stewart and Anita Raja},
  title     = {An Empirical Study of Refactorings and Technical Debt in Machine Learning Systems},
  booktitle = {Proceedings of the 43rd International Conference on Software Engineering (ICSE)},
  pages     = {238--250},
  year      = {2021},
  doi       = {10.1109/ICSE43902.2021.00033}
}

@inproceedings{rzig2023vrtesting,
  author    = {Dhia Elhaq Rzig and Nafees Iqbal and Isabella Attisano and Xue Qin and Foyzul Hassan},
  title     = {Virtual Reality (VR) Automated Testing in the Wild: a Case Study on Unity-Based VR Applications},
  booktitle = {Proceedings of the 32nd ACM SIGSOFT International Symposium on Software Testing and Analysis (ISSTA '23)},
  year      = {2023},
  doi       = {10.1145/3597926.3598134}
}

@inproceedings{bierbaum2003vrtesting,
  author    = {Aaron Bierbaum and Christopher Just and Pat Hartling and Kevin Meinert and Albert Baker and Carolina Cruz-Neira},
  title     = {Automated Testing of Virtual Reality Application Interfaces},
  booktitle = {Proceedings of the ACM Symposium on Virtual Reality Software and Technology (VRST)},
  year      = {2003},
  doi       = {10.1145/769953.769966}
}

@inproceedings{negara2013mining,
  author    = {Stas Negara and Mohsen Vakilian and Nicholas Chen and Ralph E. Johnson and Danny Dig},
  title     = {Is It Dangerous to Use Version Control Histories to Study Source Code Evolution?},
  booktitle = {Proceedings of the 27th European Conference on Object-Oriented Programming (ECOOP)},
  year      = {2013},
  pages     = {79--103},
  doi       = {10.1007/978-3-642-39038-8_4}
}

@inproceedings{ghammam2025build,
  author    = {Anwar Ghammam and Dhia Elhaq Rzig and Mohamed Almukhtar and Rania Khalsi and Foyzul Hassan and Marouane Kessentini},
  title     = {Build Code Needs Maintenance Too: A Study on Refactoring and Technical Debt in Build Systems},
  booktitle = {Proceedings of the 22nd International Conference on Mining Software Repositories (MSR 2025)},
  year      = {2025},
  pages     = {616--628},
  doi       = {10.1109/MSR66628.2025.00098}
}

@misc{vrmarket,
  title     = {Virtual Reality (VR) Market Size},
  author    = {{MarketsandMarkets}},
  year      = {2026},
  howpublished = {\url{https://www.marketsandmarkets.com/Market-Reports/reality-applications-market-458.html}},
  publisher = {MarketsandMarkets}
}

@article{zhuvrexplorer,
  title={VRExplorer: A Model-based Approach for Semi-Automated Testing of Virtual Reality Scenes},
  author={Zhu, Zhengyang and Dai, Hong-Ning and Guo, Hanyang and Liao, Zeqin and Zheng, Zibin}
}

@inproceedings{Wang2022_VRTest,
  author    = {Xiaoyin Wang},
  title     = {VRTest: An Extensible Framework for Automatic Testing of Virtual Reality Scenes},
  booktitle = {Proceedings of the 44th International Conference on Software Engineering Companion (ICSE'22 Companion)},
  year      = {2022},
  location  = {Pittsburgh, PA, USA},
  publisher = {ACM}
}

@inproceedings{Rafi2022_PredART,
  author    = {Tahmid Rafi and Xueling Zhang and Xiaoyin Wang},
  title     = {PredART: Towards Automatic Oracle Prediction of Object Placements in Augmented Reality Testing},
  booktitle = {Proceedings of the 37th IEEE/ACM International Conference on Automated Software Engineering (ASE'22)},
  year      = {2022},
  publisher = {ACM}
}

@inproceedings{wahler2016improving,
  title={Improving code maintainability: A case study on the impact of refactoring},
  author={Wahler, Michael and Drofenik, Uwe and Snipes, Will},
  booktitle={2016 IEEE International Conference on Software Maintenance and Evolution (ICSME)},
  pages={493--501},
  year={2016},
  organization={IEEE}
}

@inproceedings{oliveira2023untold,
  title={The untold story of code refactoring customizations in practice},
  author={Oliveira, Daniel and Assun{\c{c}}{\~a}o, Wesley KG and Garcia, Alessandro and Bibiano, Ana Carla and Ribeiro, M{\'a}rcio and Gheyi, Rohit and Fonseca, Baldoino},
  booktitle={2023 IEEE/ACM 45th International Conference on Software Engineering (ICSE)},
  pages={108--120},
  year={2023},
  organization={IEEE}
}

@inproceedings{palomba2017exploratory,
  title={An exploratory study on the relationship between changes and refactoring},
  author={Palomba, Fabio and Zaidman, Andy and Oliveto, Rocco and De Lucia, Andrea},
  booktitle={2017 IEEE/ACM 25th International Conference on Program Comprehension (ICPC)},
  pages={176--185},
  year={2017},
  organization={IEEE}
}

@inproceedings{peruma2022refactoring,
  title={Refactoring debt: myth or reality? an exploratory study on the relationship between technical debt and refactoring},
  author={Peruma, Anthony and AlOmar, Eman Abdullah and Newman, Christian D and Mkaouer, Mohamed Wiem and Ouni, Ali},
  booktitle={Proceedings of the 19th International Conference on Mining Software Repositories},
  pages={127--131},
  year={2022}
}

@article{traini2021software,
  title={How software refactoring impacts execution time},
  author={Traini, Luca and Di Pompeo, Daniele and Tucci, Michele and Lin, Bin and Scalabrino, Simone and Bavota, Gabriele and Lanza, Michele and Oliveto, Rocco and Cortellessa, Vittorio},
  journal={ACM Transactions on Software Engineering and Methodology (TOSEM)},
  volume={31},
  number={2},
  pages={1--23},
  year={2021},
  publisher={ACM New York, NY}
}

@inproceedings{hu2019re,
  title={Re-factoring based program repair applied to programming assignments},
  author={Hu, Yang and Ahmed, Umair Z and Mechtaev, Sergey and Leong, Ben and Roychoudhury, Abhik},
  booktitle={2019 34th IEEE/ACM International Conference on Automated Software Engineering (ASE)},
  pages={388--398},
  year={2019},
  organization={IEEE}
}

@inproceedings{meng2015does,
  title={Does automated refactoring obviate systematic editing?},
  author={Meng, Na and Hua, Lisa and Kim, Miryung and McKinley, Kathryn S},
  booktitle={2015 IEEE/ACM 37th IEEE International Conference on Software Engineering},
  volume={1},
  pages={392--402},
  year={2015},
  organization={IEEE}
}

@article{ouni2016multi,
  title={Multi-criteria code refactoring using search-based software engineering: An industrial case study},
  author={Ouni, Ali and Kessentini, Marouane and Sahraoui, Houari and Inoue, Katsuro and Deb, Kalyanmoy},
  journal={ACM Transactions on Software Engineering and Methodology (TOSEM)},
  volume={25},
  number={3},
  pages={1--53},
  year={2016},
  publisher={ACM New York, NY, USA}
}

@article{noei2025detecting,
  title={Detecting refactoring commits in machine learning python projects: A machine learning-based approach},
  author={Noei, Shayan and Li, Heng and Zou, Ying},
  journal={ACM Transactions on Software Engineering and Methodology},
  volume={34},
  number={3},
  pages={1--25},
  year={2025},
  publisher={ACM New York, NY}
}

@inproceedings{tang2021empirical,
  title={An empirical study of refactorings and technical debt in machine learning systems},
  author={Tang, Yiming and Khatchadourian, Raffi and Bagherzadeh, Mehdi and Singh, Rhia and Stewart, Ajani and Raja, Anita},
  booktitle={2021 IEEE/ACM 43rd international conference on software engineering (ICSE)},
  pages={238--250},
  year={2021},
  organization={IEEE}
}

@article{fawad2025refactoring,
  title={Refactoring Android source code smells from Android applications},
  author={Fawad, Muhammad and Rasool, Ghulam and Riaz, Muhammad Bilal},
  journal={IEEE Access},
  volume={13},
  pages={14122--14150},
  year={2025},
  publisher={IEEE}
}

@inproceedings{lin2015study,
  title={Study and refactoring of android asynchronous programming (t)},
  author={Lin, Yu and Okur, Semih and Dig, Danny},
  booktitle={2015 30th IEEE/ACM International Conference on Automated Software Engineering (ASE)},
  pages={224--235},
  year={2015},
  organization={IEEE}
}

@inproceedings{ksontini2025refactoring,
  title={Refactoring for Dockerfile Quality: A Dive into Developer Practices and Automation Potential},
  author={Ksontini, Emna and Mastouri, Meriem and Khalsi, Rania and Kessentini, Wael},
  booktitle={2025 IEEE/ACM 22nd International Conference on Mining Software Repositories (MSR)},
  pages={788--800},
  year={2025},
  organization={IEEE}
}

@inproceedings{ksontini2021refactorings,
  title={Refactorings and technical debt in docker projects: An empirical study},
  author={Ksontini, Emna and Kessentini, Marouane and Ferreira, Thiago do N and Hassan, Foyzul},
  booktitle={2021 36th IEEE/ACM International Conference on Automated Software Engineering (ASE)},
  pages={781--791},
  year={2021},
  organization={IEEE}
}

@inproceedings{lin2019unity,
  title={Why Developers Use Unity: An Empirical Study of Game and VR Projects},
  author={Lin, Y. and Dig, D. and Amann, S.},
  booktitle={Proceedings of the Mining Software Repositories Conference},
  year={2019}
}

@inproceedings{amershi2019interactive,
  title={Software Engineering Challenges of Interactive AI Systems},
  author={Amershi, S. et al.},
  booktitle={Proceedings of the International Symposium on Empirical Software Engineering and Measurement},
  year={2019}
}

@inproceedings{alomar2021refactoring,
  title={Refactoring practices in the context of modern code review: An industrial case study at Xerox},
  author={AlOmar, Eman Abdullah and AlRubaye, Hussein and Mkaouer, Mohamed Wiem and Ouni, Ali and Kessentini, Marouane},
  booktitle={2021 IEEE/ACM 43rd International Conference on Software Engineering: Software Engineering in Practice (ICSE-SEIP)},
  pages={348--357},
  year={2021},
  organization={IEEE}
}

@article{levin2019towards,
  title={Towards software analytics: Modeling maintenance activities},
  author={Levin, Stanislav and Yehudai, Amiram},
  journal={arXiv preprint arXiv:1903.04909},
  year={2019}
}

@article{usman2017taxonomies,
  title={Taxonomies in software engineering: A systematic mapping study and a revised taxonomy development method},
  author={Usman, Muhammad and Britto, Ricardo and B{\"o}rstler, J{\"u}rgen and Mendes, Emilia},
  journal={Information and Software Technology},
  volume={85},
  pages={43--59},
  year={2017},
  publisher={Elsevier}
}

@article{fincher2005making,
  title={Making sense of card sorting data},
  author={Fincher, Sally and Tenenberg, Josh},
  journal={Expert Systems},
  volume={22},
  number={3},
  pages={89--93},
  year={2005},
  publisher={Wiley Online Library}
}

@article{spencer2004card,
  title={Card sorting: a definitive guide},
  author={Spencer, Donna and Warfel, Todd},
  journal={Boxes and arrows},
  volume={2},
  number={2004},
  pages={1--23},
  year={2004}
}

@inproceedings{Herzig2013,
author = {Herzig, Kim and Zeller, Andreas},
title = {The impact of tangled code changes},
year = {2013},
isbn = {9781467329361},
publisher = {IEEE Press},
booktitle = {Proceedings of the 10th Working Conference on Mining Software Repositories},
pages = {121–130},
numpages = {10},
location = {San Francisco, CA, USA},
series = {MSR '13}
}

@inproceedings{Chen2021,
author = {Chen, Junjie and Ma, Haoyang and Zhang, Lingming},
title = {Enhanced compiler bug isolation via memoized search},
year = {2021},
isbn = {9781450367684},
publisher = {Association for Computing Machinery},
address = {New York, NY, USA},
url = {https://doi.org/10.1145/3324884.3416570},
doi = {10.1145/3324884.3416570},
booktitle = {Proceedings of the 35th IEEE/ACM International Conference on Automated Software Engineering},
pages = {78–89},
numpages = {12},
location = {Virtual Event, Australia},
series = {ASE '20}
}

@ARTICLE{Khan2022,
  author={Khan, Faizan and Chen, Boqi and Varro, Daniel and McIntosh, Shane},
  journal={IEEE Transactions on Software Engineering}, 
  title={An Empirical Study of Type-Related Defects in Python Projects}, 
  year={2022},
  volume={48},
  number={8},
  pages={3145-3158},
  doi={10.1109/TSE.2021.3082068}}

@inproceedings{Lewis2013,
author = {Lewis, Chris and Lin, Zhongpeng and Sadowski, Caitlin and Zhu, Xiaoyan and Ou, Rong and Whitehead Jr., E. James},
title = {Does bug prediction support human developers? findings from a google case study},
year = {2013},
isbn = {9781467330763},
publisher = {IEEE Press},
booktitle = {Proceedings of the 2013 International Conference on Software Engineering},
pages = {372–381},
numpages = {10},
location = {San Francisco, CA, USA},
series = {ICSE '13}
}

@inproceedings{giannaccari2025code,
  title={Code refactoring in virtual reality},
  author={Giannaccari, Mattia and Raglianti, Marco and Lanza, Michele},
  booktitle={2025 IEEE/ACM Second IDE Workshop (IDE)},
  pages={7--12},
  year={2025},
  organization={IEEE}
}

@INPROCEEDINGS{Dilhara2022,
  author={Dilhara, Malinda and Ketkar, Ameya and Sannidhi, Nikhith and Dig, Danny},
  booktitle={2022 IEEE/ACM 44th International Conference on Software Engineering (ICSE)}, 
  title={Discovering Repetitive Code Changes in Python ML Systems}, 
  year={2022},
  volume={},
  number={},
  pages={736-748},
  doi={10.1145/3510003.3510225}}

@misc{displayrefactor, title={Projection \& Display Mode Management Refactoring · BIVROST/360PlayerWindows@150f71}, url={https://github.com/BIVROST/360PlayerWindows/commit/150f711cf4070b56696920c3117b41d7b9828f76}, journal={GitHub}, language={en},month = Feb,
year = 2026 }

@misc{verefactoring,
author = {verefactoring},
title = {Is Refactoring Different in VR Software? An Empirical Study},
month = OCT,
year = 2026,
publisher = {FigShare},
url = {https://figshare.com/s/b2ea81651e9fc505b9a7},
}

@INPROCEEDINGS{Azad2023,
  author={Kalam Azad, Md Abul and Iqbal, Nafees and Hassan, Foyzul and Roy, Probir},
  booktitle={2023 IEEE/ACM 20th International Conference on Mining Software Repositories (MSR)}, 
  title={An Empirical Study of High Performance Computing (HPC) Performance Bugs}, 
  year={2023},
  volume={},
  number={},
  pages={194-206},
  doi={10.1109/MSR59073.2023.00037}}

@ARTICLE{Stana2025,
  author={Stana, Adelina Diana and Şora, Ioana},
  journal={IEEE Access}, 
  title={Refining Software Clustering: The Impact of Code Co-Changes on Architectural Reconstruction}, 
  year={2025},
  volume={13},
  number={},
  pages={132126-132145},
  doi={10.1109/ACCESS.2025.3592777}}

@INPROCEEDINGS{Nusrat2021,
  author={Nusrat, Fariha and Hassan, Foyzul and Zhong, Hao and Wang, Xiaoyin},
  booktitle={2021 IEEE/ACM 43rd International Conference on Software Engineering (ICSE)}, 
  title={How Developers Optimize Virtual Reality Applications: A Study of Optimization Commits in Open Source Unity Projects}, 
  year={2021},
  volume={},
  number={},
  pages={473-485},
  doi={10.1109/ICSE43902.2021.00052}}

@INPROCEEDINGS{Gurramkonda2025,
  author={Gurramkonda, Faraz and Chakroborty, Avishak and Maxim, Bruce and Mkaouer, Mohamed Wiem and Hassan, Foyzul},
  booktitle={2025 40th IEEE/ACM International Conference on Automated Software Engineering (ASE)}, 
  title={VRTestSniffer: Test Smell Detector for Virtual Reality (VR) Software Projects}, 
  year={2025},
  volume={},
  number={},
  pages={1956-1968},
  doi={10.1109/ASE63991.2025.00163}}

@INPROCEEDINGS{WangASE2023,
  author={Wang, Xiaoyin and Rafi, Tahmid and Meng, Na},
  booktitle={2023 38th IEEE/ACM International Conference on Automated Software Engineering (ASE)}, 
  title={VRGuide: Efficient Testing of Virtual Reality Scenes via Dynamic Cut Coverage}, 
  year={2023},
  volume={},
  number={},
  pages={951-962},
  doi={10.1109/ASE56229.2023.00197}}

@INPROCEEDINGS{Zhu2025,
  author={Zhu, Zhengyang and Dai, Hong-Ning and Guo, Hanyang and Liao, Zeqin and Zheng, Zibin},
  booktitle={2025 40th IEEE/ACM International Conference on Automated Software Engineering (ASE)}, 
  title={VRExplorer: A Model-based Approach for Semi-Automated Testing of Virtual Reality Scenes}, 
  year={2025},
  volume={},
  number={},
  pages={482-494},
  doi={10.1109/ASE63991.2025.00047}}

@ARTICLE{Murphy-Hill2012,
  author={Murphy-Hill, Emerson and Parnin, Chris and Black, Andrew P.},
  journal={IEEE Transactions on Software Engineering}, 
  title={How We Refactor, and How We Know It}, 
  year={2012},
  volume={38},
  number={1},
  pages={5-18},
  doi={10.1109/TSE.2011.41}}

@inproceedings{Tsantalis2013,
author = {Tsantalis, Nikolaos and Guana, Victor and Stroulia, Eleni and Hindle, Abram},
title = {A multidimensional empirical study on refactoring activity},
year = {2013},
publisher = {IBM Corp.},
address = {USA},
booktitle = {Proceedings of the 2013 Conference of the Center for Advanced Studies on Collaborative Research},
pages = {132–146},
numpages = {15},
location = {Ontario, Canada},
series = {CASCON '13}
}

@article{HABCHI2021110964,
title = {Android code smells: From introduction to refactoring},
journal = {Journal of Systems and Software},
volume = {177},
pages = {110964},
year = {2021},
issn = {0164-1212},
doi = {https://doi.org/10.1016/j.jss.2021.110964},
url = {https://www.sciencedirect.com/science/article/pii/S0164121221000613},
author = {Sarra Habchi and Naouel Moha and Romain Rouvoy}
}

@INPROCEEDINGS{Ksontini2025,
  author={Ksontini, Emna and Mastouri, Meriem and Khalsi, Rania and Kessentini, Wael},
  booktitle={2025 IEEE/ACM 22nd International Conference on Mining Software Repositories (MSR)}, 
  title={Refactoring for Dockerfile Quality: A Dive into Developer Practices and Automation Potential}, 
  year={2025},
  volume={},
  number={},
  pages={788-800},
  doi={10.1109/MSR66628.2025.00116}}

@article{LiASE2026,
author = {Li, Meiziniu and Li, Dongze and Liu, Jianmeng and Cao, Jialun and Tian, Yongqiang and Cheung, Shing-Chi},
title = {Enhancing Differential Testing with LLMs for Testing Deep Learning Libraries},
year = {2026},
issue_date = {April 2026},
publisher = {Association for Computing Machinery},
address = {New York, NY, USA},
volume = {35},
number = {4},
issn = {1049-331X},
url = {https://doi.org/10.1145/3735637},
doi = {10.1145/3735637},
journal = {ACM Trans. Softw. Eng. Methodol.},
month = mar,
articleno = {88},
numpages = {39}
}

@inproceedings{Dintzner2016,
author = {Dintzner, Nicolas and van Deursen, Arie and Pinzger, Martin},
title = {FEVER: extracting feature-oriented changes from commits},
year = {2016},
isbn = {9781450341868},
publisher = {Association for Computing Machinery},
address = {New York, NY, USA},
url = {https://doi.org/10.1145/2901739.2901755},
doi = {10.1145/2901739.2901755},
booktitle = {Proceedings of the 13th International Conference on Mining Software Repositories},
pages = {85–96},
numpages = {12},
location = {Austin, Texas},
series = {MSR '16}
}

@inproceedings{borrelli2020unity,
  author    = {Antonio Borrelli and
               Vittoria Nardone and
               Giuseppe A. Di Lucca and
               Gerardo Canfora and
               Massimiliano Di Penta},
  title     = {Detecting Video Game-Specific Bad Smells
               in Unity Projects},
  booktitle = {Proceedings of the 17th International
               Conference on Mining Software Repositories
               (MSR)},
  year      = {2020},
  pages     = {198--208},
  doi       = {10.1145/3379597.3387454}
}

@INPROCEEDINGS{Silva2017,
  author={Silva, Danilo and Valente, Marco Tulio},
  booktitle={2017 IEEE/ACM 14th International Conference on Mining Software Repositories (MSR)}, 
  title={RefDiff: Detecting Refactorings in Version Histories}, 
  year={2017},
  volume={},
  number={},
  pages={269-279},
  doi={10.1109/MSR.2017.14}}

@ARTICLE{Moghadam2021,
  author={Moghadam, Iman Hemati and {\'O} Cinn{\'e}ide, Mel and Zarepour, Faezeh and Jahanmir, Mohamad Aref},
  journal={IEEE Access}, 
  title={RefDetect: A Multi-Language Refactoring Detection Tool Based on String Alignment}, 
  year={2021},
  volume={9},
  number={},
  pages={86698-86727},
  doi={10.1109/ACCESS.2021.3086689}}

@misc{sysmagine2026semanticdiff,
  author       = {{Sysmagine GmbH}},
  title        = {{SemanticDiff}: Language-Aware Diff for
                  Visual Studio Code and GitHub},
  howpublished = {\url{https://semanticdiff.com/}},
  note         = {Accessed: July 19, 2026},
  year         = {2026}
}

@mastersthesis{kalaitzidis2025refactoringminer,
  author = {Dimitrios Kalaitzidis},
  title  = {RefactoringMiner Extension},
  school = {University of Macedonia},
  year   = {2025},
  url    = {https://dspace.lib.uom.gr/bitstream/2159/33815/1/KalaitzidisDimitriosMsc2025.pdf}
}

@misc{sonarsource2025sonarqube,
  author       = {{SonarSource}},
  title        = {{SonarQube Server}: Understanding and Resolving Issues},
  year         = {2025},
  howpublished = {\url{https://docs.sonarsource.com/sonarqube-server/2025.5/user-guide/issues/solution-overview}},
  note         = {Version 2025.5 documentation, accessed October 19, 2026}
}

@misc{jetbrains2026inspectcode,
  author       = {{JetBrains}},
  title        = {{InspectCode}: Command-Line Code Inspection},
  year         = {2026},
  howpublished = {\url{https://www.jetbrains.com/help/resharper/InspectCode.html}},
  note         = {Accessed: June 19, 2026}
}

@article{nardone2023videogame,
  author  = {Vittoria Nardone and
             Biruk Muse and
             Mouna Abidi and
             Foutse Khomh and
             Massimiliano Di Penta},
  title   = {Video Game Bad Smells: What They Are
             and How Developers Perceive Them},
  journal = {ACM Transactions on Software
             Engineering and Methodology},
  volume  = {32},
  number  = {4},
  articleno = {88},
  year    = {2023},
  doi     = {10.1145/3563214}
}

@inproceedings{bosco2023unitylint,
  author    = {Matteo Bosco and
               Pasquale Cavoto and
               Augusto Ungolo and
               Biruk Asmare Muse and
               Foutse Khomh and
               Vittoria Nardone and
               Massimiliano Di Penta},
  title     = {UnityLint: A Bad Smell Detector for Unity},
  booktitle = {Proceedings of the 31st IEEE/ACM
               International Conference on
               Program Comprehension (ICPC)},
  pages     = {186--190},
  year      = {2023},
  doi       = {10.1109/ICPC58990.2023.00033}
}

@article{xr_development_2026,
  title   = {XR application development in the 21st century:
             a survey spanning two decades of XR developers,
             applications, and challenges},
  journal = {Frontiers in Virtual Reality},
  year    = {2026},
  doi     = {10.3389/frvir.2026.1769463}
}

\end{document}